\documentstyle[prl,aps,multicol,psfig,epsfig]{revtex}

\begin{document}
\draft
\begin{multicols}{2}
{\bf Pudalov {\em et al}. Reply}. In a recent Letter \cite{aniso}
we studied the magnetoresistance (MR) in strong in-plane fields
for a number of Si-MOSFET samples with different mobilities. We
found that the field of the MR saturation $B_{\rm sat}$ is
strongly sample-dependent, being different by a factor of up to
two for different samples at a given  density of mobile electrons.
In subsequent studies \cite{cooldown}, a similar observation  was
reported for a single sample, cooled down at different gate
voltage values. Based on these results,
 we highlighted in Ref.~\cite{aniso} the role of sample-specific
localized states (``disorder'') in the strong-field MR, and
concluded that the field at which the MR saturates does not
reflect spin-polarization of mobile carriers solely.

In the  Comment to our paper \cite{DGcomment},
Dolgopolov and Gold (DG) attempt to explain the experimentally
observed non-universality (i.e. the sample dependence) of the
$B_{\rm sat}$ field. For the two-dimensional (2D) case, they
apply a two-fluid model, which is known for the 3D Anderson-Mott transition,
with the aim to calculate a disorder-dependent
field of complete spin polarization.
DG do not calculate the MR, but implicitly identify $B_c$ with $B_{\rm sat}$,
thus using the assumption which was questioned in Ref.~\cite{aniso}.

The  density of electrons $n$, which is plotted on the horizontal
axes of Fig.~2 of Ref.~\cite{aniso}, is deduced from the period
of Shubnikov-de Haas  oscillations, as stated in our paper. The
oscillations in the 2D case have a frequency equal to $n_{\rm
SdH}/n_h$ where $n_h$ is the density of flux quanta, and $n_{\rm
SdH}$ is the  density of those electrons which
participate in cyclotron motion and
occupy the Landau levels.

In order to explain our observation of the disorder-dependent
$B_{\rm sat}$,
DG suggest to deduce the
density of {\em extended states} $n_{\rm ext}$  by  subtracting
the density of
singly occupied localized states $n_{so}$ from the density
$n=n_{\rm SdH}$.
The relation $n_{\rm ext} = n_{\rm SdH} - n_{so}$
were automatically fulfilled  if all electrons
(localized and extended) would participate in the SdH effect.
However, it can be questioned whether this is the case,
and the  DG model does not answer this question.
The ``reduced'' quantity $n_{\rm ext}$ is further used in order
to obtain a reduced field $B_c$
of the complete spin polarization of extended states.
To calculate the reduced $B_c$ value, DG  take into account a
constant density of states $\rho_F$. But in their model,
$\rho_F$ has to be reduced for lower lying states  and the shift
in $B_c$ might have opposite sign! Definitely, the behavior of
localized states in magnetic field in the Hubbard model with
strong on-site interaction requires a more thorough consideration.

We note that in order to explain the observed changes in
$B_{\rm sat}$ of up to 2\,Tesla (see e.g. Fig.~2 of Ref.~\cite{aniso}
 and Fig.~4a of Ref.~\cite{cooldown}) in the DG model, the
density of extended states must be reduced by
 $\approx 1\times 10^{11}$cm$^{-2}$.
In a typical situation where the total density of electrons in the sample
(calculated from the capacitance) is  $1\times 10^{11}$cm$^{-2}$
(see e.g. Fig.~1 of Ref.~\cite{cooldown}) and equals to within
5\% to the measured $n_{\rm SdH}$, such reduction would lead
to practically zero density of mobile electrons, enormously
reduced density of states and effective mass;
all these effects are not observed and seem very unlikely.

Another point concerns the second paragraph of the Comment.
The authors misguide the
readers in  prescribing a statement to our paper  about an
interrelation between the Hall voltage and the  SdH period, which
is not used there. More over, just opposite results have been
observed and reported by us in Ref.~\cite{JETPLHall}.
Thus, DG use arguments which
do not belong to our paper and which contradict our view. Also
misleading is that DG ignore the data for two samples (rather
than one) in Fig.~2 of Ref.~\cite{aniso} which disagrees with
their model, and ignore the linear dependence of the offset
density $n_d$ vs inverse sample mobility (which can be noticed
from the data reported in Ref.~\cite{aniso}). The latter suggests
that an ``ideal sample'' (with infinite mobility) would show
$B_{\rm sat}$ substantially larger than the calculated spin
polarization field $B_{c0}$.

Further, the approach of DG is incomplete.  The main equation of
proportionality   between
$B_{c0}$ and the density means that DG assume a
constant 2D density of states to be valid for  the strongly
interacting 2D liquid at $r_s \sim 10$. There is neither
theoretical nor experimental cause to believe that
the relation between the $B_{c0}$ and $E_F$ can
be so simple. The problem of
the spin-polarization in fields $\sim E_F$  is a large energy
problem, beyond the frameworks of the Fermi-liquid concept.

To conclude, we agree with the authors of the Comment in the sense
that the localized electrons and local moments associated with
localized states in 2D systems play an important role in
transport at low densities. The experimental evidence for this is
one of the main results of our paper \cite{aniso}. However, we
disagree with the oversimplified proposed  model as
it is inconsistent with the experimental
results.

Authors acknowledge support by FWF Austria, INTAS, NATO, NSF,
RFBR, and the programs
 ``Physics of nanostructures'',
``Quantum and non-linear processes'', ``Integration of high
education and academic research'', ``Quantum computing and
telecommunications'', and ``The State support of leading
scientific schools''.
\vspace{0.2in}

 V.\ M.\ Pudalov$^{a,b}$, G.\ Brunthaler$^c$, A.\
Prinz$^c$, and
G.\ Bauer$^c$\\
$^a$ P.\ N.\ Lebedev Physics Institute, Moscow, Russia.\\
$^b$ Department of Physics and Astronomy, Rutgers University, NJ,
USA\\
$^c$ Institut f\"{u}r Halbleiterphysik, Johannes Kepler
Universt\"{a}t, Linz, Austria.

\end{multicols}

\vspace{-0.1in}


\begin{references}
\vspace{-0.4in}

\bibitem{aniso}V.\ M.\ Pudalov, G.\ Brunthaler, A.\ Prinz, and
G.\ Bauer, Phys. Rev. Lett. {\bf 88} 176401 (2002).

\bibitem{cooldown}V.\ M.\ Pudalov, M.\ E.\ Gershenson, H.\ Kojima,
cond-mat/0201001.
\bibitem{DGcomment}V.\ T.\ Dolgopolov, A.\ V.\ Gold,
cond-mat/0203276.

\bibitem{JETPLHall}V.\ M.\ Pudalov, G.\ Brunthaler, A.\ Prinz, and
G.\ Bauer, JETP Lett. {\bf 70}, 48 (1999).
\end{references}
\end{document}